\documentclass[12pt]{article}
\usepackage{graphicx}
\usepackage{times}

\usepackage{epsf}
\def\beq{\begin{equation}}
\def\eeq{\end{equation}}
\def\bea{\begin{eqnarray}}
\def\eea{\end{eqnarray}}

\def\vel{\left|}
\def\ver{\right|}
\def\nnb{\nonumber}

\def\rar{\rightarrow}
\def\nnb{\nonumber}

\def\ba{\begin{array}}
\def\ea{\end{array}}
\def\bea{\begin{eqnarray}}
\def\eea{\end{eqnarray}}

\def\vel{\left|}
\def\ver{\right|}
\def\nnb{\nonumber}

\def\rar{\rightarrow}
\def\nnb{\nonumber}

\def\lla{\left<}
\def\rra{\right>}

\begin{document}

\title{ {\Large {\bf Double-Lepton Polarization
Asymmetries in $B \rar K_1 l^+ l^- $ Decay in Universal Extra
Dimension Model }}}
\author{ {\small B. B.  \c{S}irvanl{\i}}\\
{\small Gazi University, Faculty of Arts and Science, Department of Physics} \\
{\small 06100, Teknikokullar Ankara, Turkey}}

\thispagestyle{empty}
\maketitle

\begin{abstract}
Double-lepton polarization asymmetries for the exclusive decay $B
\rar K_1 l^+ l^- $  in the Universal Extra Dimension (UED) Model
is studied. It is obtained that double-lepton polarization
asymmetries are very sensitive to the UED model parameters.
Experimental measurements of double lepton polarizations can give
valuable information on the physics beyond the Standard Model
(SM).
\end{abstract}
~~~PACS number(s):12.60.--i, 13.20.--v, 13.20.He

%
\clearpage

\section{Introduction \label{s1}}
The rare B-meson decays pointed out by the flavor-changing neutral
currents (FCNC) have been significant channels for acquiring
knowledge on the SM parameter and analyzing the new physics
predictions. Rare B meson decays are not allowed at the tree level
in the SM and seem at loop level. By rare B decays, one generally
comprehend Cabibbo-suppressed $b \rar u$ transitions or
flavour-changing neutral currents (FCNC) $b \rar s$ or $b \rar d$.
So rare decays are significant testing basic of the SM and take an
important part in the search for new physics. The examinations of
different FCNC processes can be used to determine different
fundamental parameters of SM like elements of the
Cabibbo-Kobayashi-Maskawa (CKM) matrix , various decay constants
etc. Between testing SM the FCNC processes can be very important
for discovering indirect effects of possible TeV scale extensions
of SM. Therefore,we examine $b \rar q (q=d,s)$ transitions in
terms of an effective Hamiltonian. For observing to the new
physics in these decays, there are two different ways. First of
all, the differences in the Wilson coefficients form the ones
existing in the SM. And the second one the new operator in the
effective Hamiltonian which are absent in the SM. All decay
channels of B meson include many physically quantities which are
very useful testing for the SM and investigating for new physics
beyond the SM. Exclusive processes such as $B\rightarrow K
(K^{*})l^{+}l^{-}$ and $B\rightarrow \gamma l^{+}l^{-}$ decays
\cite {Ali,Aliev1,Kruger,Rai,sirvanli1} have been studied
extensive in literature . Colangelo et al. have studied
$B\rightarrow K (K^{*})l^{+}l^{-}$ decays in framework of one
Universal Extra Dimension model (ACD),proposed in the Ref. \cite
{Colangelo} and analyzed the branching ratio and forward-backward
asymmetry . In meanwhile, in the Ref. \cite {ahmed} the single
lepton polarizations is studied for $\mu$ for the $B \rar K_1 l^+
l^- $ decay in UED model. The Branching ratios (BR) of the
Semileptonic decays $\mathcal{B}(B\rightarrow K^{*}l^{+}l^{-}) =
7.8\pm1.2\times10^{-7}$ \cite {Aubert} and
$\mathcal{B}(B\rightarrow K l^{+}l^{-}) = 5.5\pm0.02\times10^{-7}$
\cite {Abe} have been measured by BELLE \cite {Abe} and BaBar
\cite {Aubert} collaborations. It is noted that the measurement of
the polarization of the $b \rar s$ decay can provide important
information about more observables. Some of the single lepton
polarization asymmetries can be too small to be observed. Since it
might not provide number of observables for control the structure
of the effective Hamiltonian,we calculate to double lepton
polarization for more observables \cite{london}. Among the
different models of physics beyond the SM, extra dimensions is
very interesting models. Since the extra dimension model contain
of gravity, they give to clue on the hierarchy problem and a
connection with string theory. The model of Appelquist, Cheng and
Dobrescu (ACD) \cite{appel,Buras1,Lü} with one universal extra
dimension (UED), where all the SM particles can propagate in the
extra dimension. Compactification of the extra dimension leads to
Kaluza-Klein model in the four-dimension. In the extra dimension
model, we have extra free parameter is $1/R$,which is inverse of
the compactification radius. With the aid of $1/R$, we can
determined all the masses of the KK particles and their
interactions with SM particles. In the meanwhile, If we have not
tree level contribution of KK states to the low energy processes,
KK parity is conservation in ACD model at scale $\mu\ll1/R$.

In this work, we study the double-lepton polarization asymmetries
for the $B \rar K_1 l^+ l^- $ decay in the UED model. In section
2, we shortly examine ACD model. In section 3, we obtain matrix
element for the $B \rar K_1 l^+ l^- $ decay. In section 4, Double
lepton polarization for the $B \rar K_1 l^+ l^- $ decay are
calculated. Section 5 is devoted to the numerical analysis and
discussion of our results.

\section{$B \rar K_1 l^+ l^- $ Decay in ACD Model \label{s2}}

Before calculation of the double lepton polarizations few words
about the ACD model. This model is the minimal extension of the SM
to the $4+\delta$ dimensions. We consider simple case which is
$\delta=1$. In the universe, we have 3 space + 1 time dimensions
and one possibility is the propagation of gravity in whole
ordinary plus extra dimensional universe. The five-dimensional ACD
model with a single UED uses orbifold compactification, the fifth
dimension $y$ that is compactified in a circle of radius $R$, with
points $y = 0$ and $y = \pi R$ that are fixed points of the
orbifolds \cite{Buras1,Buras2,Aliev2,ahmed}. The Lagrangian in ACD
model can be written as:

\bea\mathcal{L}=\int d^{4}x dy
\{\mathcal{L}_{A}+\mathcal{L}_{H}+\mathcal{L}_{F}+\mathcal{L}_{Y}\}\nnb
\eea

where

\bea\mathcal{L}_{A}&=&-\frac{1}{4}W^{MNa}W_{MN}^{a}-\frac{1}{4}B^{MN}B_{MN}
\nnb\\\mathcal{L}_{H}&=&(\mathcal{D}^{M}\phi)^{\dagger}\mathcal{D}_{M}\phi-V(\phi)\nnb\\
\mathcal{L}_{F}&=&\overline{\mathcal{Q}}(i
\Gamma^{M}\mathcal{D}_{M})\mathcal{Q}+\overline{u}(i
\Gamma^{M}\mathcal{D}_{M})u+\overline{\mathcal{D}}(i
\Gamma^{M}\mathcal{D}_{M})\mathcal{D}\nnb\\
\mathcal{L}_{Y}&=&
-\overline{\mathcal{Q}}\widetilde{Y}_{u}\phi^{c}u-\overline{\mathcal{Q}}\widetilde{Y}_{d}\phi\mathcal{D}+h.c.
.\nnb \eea

where $M$ and $N$ are the five-dimensional Lorentz indices which
can run from $0,1,2,3,5$.
$W_{MN}^{a}=\partial_{M}W_{N}^{a}-\partial_{N}W_{M}^{a}+\widetilde{g}\varepsilon^{abc}W_{M}^{b}W_{N}^{c}$
are the field strength tensor for the $SU(2)_{L}$ electroweak
group, $B_{MN}=\partial_{M}B_{N}-\partial_{N}B_{M}$ are that of
the $U(1)$ group.
$\mathcal{D}_{M}=\partial_{M}-i\widetilde{g}W_{M}^{a}T^{a}-i\widetilde{g}^{'}B_{M}Y$
is the covariant derivative, where $\widetilde{g}$ and
$\widetilde{g}^{'}$ are the five-dimensional gauge couplings for
the $SU(2)_{L}$ and $U(1)$ groups. $\Gamma^{M}$ are
five-dimensional matrices which is $\Gamma^{\mu}=\gamma^{\mu}$ ,
$\mu=0,1,2,3$ and $\Gamma^{5}=i\gamma^{5}$. $F(x_{t},y)$ is the
periodic function of $y$ which is $1/R$. It can be written as
follow:

\bea F(x_{t},y) =F_{0}(x_{t})+\sum_{n=1}^{+\infty}
F_{n}(x_{t},x_{n}) \nnb \eea

where $x_{t}=\frac{m_{t}^{2}}{m_{w}^{2}}$,
$x_{n}=\frac{m_{n}^{2}}{m_{w}^{2}}$ and $m_{n}=n/R$. These
function can be found in [10,15].

\section{Effective Hamiltonian for $B \rar K_1 l^+ l^- $  Decay \label{s3}}

At quark level, the exclusive $B \rar K_1 l^+ l^- $  decay is
described by $b\rightarrow s l^{+}l^{-}$ transition governed by
effective Hamiltonian:

\bea \label{effH} {\cal H}_{eff}&=& -4 \frac{G_{F}}{\sqrt{2}}
V_{tb} V_{ts}^* \sum_{i=1}^{10} C_{i}(\mu)O_{i}(\mu)   \eea where
$O_{i}$'s are local quark operators and $C_{i}$'s are Wilson
coefficients. $G_{F}$ is the Fermi constant and $V_{ij}$ are
elements of the Cabibbo-Kobayashi-Maskawa (CKM) matrix element for
$B \rar K_1 l^+ l^- $ decay is obtained by $b\rightarrow s
l^{+}l^{-}$ sandwiching transition amplitude between initial and
final meson states. Using effective Hamiltonian the matrix element
of the $B \rar K_1 l^+ l^- $ decay which can be written as
follows:

\bea\mathcal{M}&=&\frac{G_{F}\alpha}{2\sqrt{2}\pi}V_{tb}
V_{ts}^*\Bigg\{-2m_{b}C^{eff}_{7}\overline{s}i\sigma_{\mu\nu}q^{\nu}(1+\gamma_{5})b\overline{l}\gamma^{\mu}l\nnb \\
&+&C^{eff}_{9}\overline{s}\gamma_{\mu}(1-\gamma_{5})b\overline{l}\gamma^{\mu}l+
C_{10}\overline{s}\gamma_{\mu}(1-\gamma_{5})b\overline{l}\gamma^{\mu}\gamma_{5}l\Bigg\}\eea
where $s=q^{2}$, q is the momentum transfer,
$q=p_{1}+p_{2}=p_{B}-p_{K_{1}}$. Here, $p_{1}$, $p_{2}$, $p_{B}$
and $p_{K_{1}}$ are the four-momenta of the leptons, $B$ meson and
$K_{1}$ meson respectively. Already the free quark decay amplitude
$\mathcal{M}$ contains certain long-distance effects which usually
are absorbed into a redefinition of the Wilson coefficient. These
coefficients in UED are calculated by Ref.\cite{Buras1} and
\cite{Buras2} which can be written as
follows, \bea C^{0}_{7}(\mu_{w})&=& -\frac{1}{2}D^{'}(x_{t},1/R),\nnb\\
C_{9}(\mu)&=& P^{NDR}_{0}+\frac{Y(x_{t},1/R)}{sin^{2}\theta_{w}}-4Z(x_{t},1/R)+P_{E}E(x_{t},1/R),\nnb\\
C_{10}&=& -\frac{Y(x_{t},1/R)}{sin^{2}\theta_{w}}\eea where
$P^{NDR}_{0}=2.60\pm0.25$ and referring to leading log
approximation. Explicit expression the functions of the detail
$D^{'}(x_{t},1/R),Y(x_{t},1/R)$ and $Z(x_{t},1/R)$ are calculated
in Ref.\cite{Buras1,Buras2,Colangelo}. From Eq.(2) it follows
that, for obtaining matrix element for the $B \rar K_1 l^+ l^- $
decay we need to know following matrix elements $ \lla
K_{1}(k,\varepsilon) \vel \bar s \gamma_\mu (\gamma_5) b \ver B(p)
\rra$ and $\lla K_{1}(k,\varepsilon) \vel \bar s i \sigma_{\mu\nu}
q^{\nu} b \ver B(p) \rra$. These matrix elements in terms of form
factors are parametrized  as

\begin{eqnarray}
  \lla K_{1}(k,\varepsilon) \vel \bar s \gamma_\mu
b \ver B(p) \rra  &=& i\varepsilon^{*}_{\mu}(m_{B}+m_{K_{1}})V_{1}(s)-(p+k)_{\mu}(\varepsilon^{*}.q)\frac{V_{2}(s)}{m_{B}+m_{K_{1}}}\nnb\\
&-&q_{\mu}(\varepsilon.q)\frac{2m_{K_{1}}}{s}[V_{3}(s)-V_{0}(s)]~, \\
  \lla K_{1}(k,\varepsilon) \vel \bar s
\gamma_\mu\gamma_5 b \ver B(p) \rra &=&
\frac{2i\epsilon_{\mu\nu\alpha\beta}}{m_{B}+m_{K_{1}}}\varepsilon^{*\nu}p^{\alpha}k^{\beta}A(s)
\end{eqnarray}

\begin{eqnarray}
  \lla K_{1}(k,\varepsilon) \vel \bar s i \sigma_{\mu\nu} q^{\nu}
b \ver B(p) \rra
&=&\Bigg[(m_{B}^{2}-m_{K_{1}}^{2})\varepsilon_{\mu}-(\varepsilon.q)(p+k)_{\mu}\Bigg]F_{2}(s)\nnb\\
&+& (\varepsilon^{*}.q)
\Bigg[q_{\mu}-\frac{s}{m_{B}^{2}-m_{K_{1}}^{2}}(p+k)_{\mu}\Bigg]F_{3}(s)~, \\
  \lla K_{1}(k,\varepsilon) \vel \bar s i
\sigma_{\mu\nu} q^{\nu}\gamma_5 b \ver B(p) \rra &=& -i
\epsilon_{\mu\nu\alpha\beta}\varepsilon^{*\nu}k^{\beta}F_{1}(s)
\end{eqnarray}
where $\varepsilon$ is the polarization vector of the $K_{1}$
meson. The form factors entering Eq.(4) and (5) are estimated in
\cite{Gilani,Dominguez}.

\begin{eqnarray}
 V_{1}(s) &=& \frac{V_{1}(0)}{(1-s/m^{2}_{B^{*}_{A}})(1-s/m^{'2}_{B^{*}_{A}})}\Bigg(1-\frac{s}{m_{B}^{2}-m_{K_{1}}^{2}}\Bigg) \\
  V_{2}(s) &=& \frac{\tilde{V_{2}}(0)}{(1-s/m^{2}_{B^{*}_{A}})(1-s/m^{'2}_{B^{*}_{A}})}-
\frac{2m_{K_{1}}}{m_{B}-m_{K_{1}}}\frac{V_{0}(0)}{(1-s/m^{2}_{B})(1-s/m^{'2}_{B})} \\
  V_{3}(s) &=& \frac{m_{B}+m_{K_{1}}}{2m_{K_{1}}}V_{1}(s)-
\frac{m_{B}-m_{K_{1}}}{2m_{K_{1}}}V_{2}(s) \\
  A(s) &=& \frac{A(0)}{(1-s/m^{2}_{B})(1-s/m^{'2}_{B})}
\end{eqnarray}

We can also define to the other matrix elements of the $B \rar K_1
l^+ l^- $ decay in terms of penguin form factors. Using the Ward
identities following relationship between form factors, we get

\begin{eqnarray}
  F_{1}(s) &=& -\frac{(m_{b}-m_{s})}{(m_{B}+m_{K_{1}})}2A(s) \\
  F_{2}(s) &=& -\frac{(m_{b}+m_{s})}{(m_{B}-m_{K_{1}})}V_{1}(s) \\
  F_{3}(s) &=& \frac{2m_{K_{1}}}{s}(m_{b}+m_{s})[V_{3}(s)-V_{0}(s)]
\end{eqnarray}

In order to avoid the kinematical singularity in the matrix
element at $s=0$ we demand $F_{1}(0)=2F_{2}(0)$. The corresponding
values at $s=0$ are given by
\cite{ahmed,paracha,Gilani,Dominguez},

\begin{eqnarray}
  A(0) &=& -(0.52\pm0.05) \nnb\\
  V_{1}(0) &=& -(0.24\pm0.02) \nnb\\
  \tilde{V_{1}}(0) &=& -(0.39\pm0.03) \nnb\\
  V_{0}(0) &=& -(0.29\pm0.04) \nnb\\
  A_{1}(0) &=& (0.23\pm0.02) \nnb\\
  \tilde{A_{2}}(0) &=& (0.33\pm0.05)
\end{eqnarray}

Using Eq.(4),(5)(6) and (7) for the matrix element of the $B \rar
K_1 l^+ l^- $ decay we set,

\bea\mathcal{M}&=&\frac{G_{F}\alpha}{2\sqrt{2}\pi}V_{tb} V_{ts}^*
m_{B}
\Bigg\{\Bigg[A(\hat{s})\epsilon_{\mu\nu\alpha\beta}\epsilon^{*\nu}p^{\alpha}_{B}p^{\beta}_{K_{1}}-i
B(\hat{s})\epsilon^{*}_{\mu}+i
C(\hat{s})(\epsilon^{*}.p_{B})(p_{B}+p_{K_{1}})_{\mu}\nnb\\&+&i
D(\hat{s})(\epsilon^{*}.p_{B})q_{\mu}\Bigg](\bar{l}\gamma^{\mu}l)
+\Bigg[E(\hat{s})\epsilon_{\mu\nu\alpha\beta}\epsilon^{*\nu}p^{\alpha}_{B}p^{\beta}_{K_{1}}-i
F(\hat{s})\epsilon^{*}_{\mu}\nnb\\ &+&i
G(\hat{s})(\epsilon^{*}.p_{B})(p_{B}+p_{K_{1}})_{\mu}+i
H(\hat{s})(\epsilon^{*}.p_{B})q_{\mu}\Bigg](\bar{l}\gamma^{\mu}\gamma^{5}l)\Bigg\}\eea

where

\begin{eqnarray}
  A(\hat{s}) &=&
  -\frac{2A(\hat{s})}{1+\frac{m_{K_{1}}}{m_{B}}}C^{eff}_{9}(\hat{s})+\frac{2m_{b}}{m_{B}\hat{s}}C^{eff}_{7}F_{1}(\hat{s})
  \nnb\\
   B(\hat{s}) &=& (1+\frac{m_{K_{1}}}{m_{B}})\Bigg[C^{eff}_{9}(\hat{s})V_{1}(\hat{s})+\frac{2m_{b}}{m_{B}\hat{s}}
   C^{eff}_{7}(1-\frac{m_{K_{1}}}{m_{B}})\Bigg]  \nnb\\
   C(\hat{s}) &=& \frac{1}{(1-(\frac{m_{K_{1}}}{m_{B}})^2)}\Bigg\{C^{eff}_{9}(\hat{s})V_{2}(\hat{s})+\frac{2m_{b}}{m_{B}}
   C^{eff}_{7}\Bigg[F_{3}(\hat{s})+\frac{1-(\frac{m_{K_{1}}}{m_{B}})^2}{\hat{s}}F_{2}(\hat{s})\Bigg] \Bigg\} \nnb\\
   D(\hat{s}) &=& \frac{1}{\hat{s}}\Bigg[\Bigg( C^{eff}_{9}(\hat{s})(1+\frac{m_{K_{1}}}{m_{B}})V_{1}(\hat{s})
   -(1-\frac{m_{K_{1}}}{m_{B}})V_{2}(\hat{s})-2\frac{m_{K_{1}}}{m_{B}}V_{0}(\hat{s})\Bigg)-2\frac{2m_{b}}{m_{B}} C^{eff}_{7}F_{3}(\hat{s})\Bigg]  \nnb\\
   E(\hat{s}) &=& -\frac{2A(\hat{s})}{1+\frac{m_{K_{1}}}{m_{B}}}C_{10}  \nnb\\
   F(\hat{s}) &=& (1+\frac{m_{K_{1}}}{m_{B}}) C_{10}V_{1}(\hat{s}) \nnb\\
   G(\hat{s}) &=& \frac{1}{(1+\frac{m_{K_{1}}}{m_{B}})} C_{10}V_{2}(\hat{s})  \nnb\\
   H(\hat{s}) &=& \frac{1}{\hat{s}}\Bigg[C_{10}(\hat{s})(1+\frac{m_{K_{1}}}{m_{B}})V_{1}(\hat{s})-
   (1-\frac{m_{K_{1}}}{m_{B}})V_{2}(\hat{s})-2\frac{m_{K_{1}}}{m_{B}}V_{0}(\hat{s})\Bigg]
\end{eqnarray}

Having the explicit expression for the matrix element for the $B
\rar K_1 l^+ l^- $ decay, the next task is the calculation its
differential decay rate. In the center of mass frame (CM) of the
dileptons $l^{+}l^{-}$, where we take $z=cos\theta$ and $\theta$
is the angle between the momentum of the $B$ meson and that of
$l^{-}$, differential decay width is found to belike follows,

\bea\frac{d\Gamma}{d\hat{s}}(B \rar K_1 l^+
l^-)=\frac{G_{F}^{2}\alpha^{2}|V_{tb} V_{ts}^*|^{2}}{8
m_{B}^{4}\pi^{2}}\Delta\eea

where $\lambda=r^2+(-1+\hat{s})^2-2r(1+\hat{s})$ with
$\hat{s}=q^2/m_{B}^2$ and $r=m_{l}^{2}/m_{B}^2$ and
$\hat{m_{l}}=m_{l}/m_{B}$. $s=q^2$ is the dilepton invariant mass.
The function $\Delta$ is defined as follows:

\bea\Delta&=&\frac{2}{3}m_{B}^{2}\Bigg\{2 m_{B}^{4}
(2\hat{m_{l}}^2+\hat{s})\lambda
|A|^2+\frac{1}{r\hat{s}}(2\hat{m_{l}}^2+\hat{s})(r^2+(-1+\hat{s})^2+2
r (-1+5\hat{s}))|B|^2 \nnb\\&+&
\frac{1}{r\hat{s}}m_{B}^{4}(2\hat{m_{l}}^2+\hat{s})\lambda^{2}|C|^2-2
m_{B}^{4}(4\hat{m_{l}}^2-\hat{s})\lambda|E|^2\nnb\\&+&\frac{1}{r\hat{s}}\Bigg[\hat{s}
\Bigg(r^2+(-1+\hat{s})^2+2 r
(-1+5\hat{s})\Bigg)+2\hat{m_{l}}^2\Bigg(r^2+(-1+\hat{s})^2-2 r
(1+13\hat{s}\Bigg)\Bigg]|F|^2\nnb\\&+&
\frac{1}{r\hat{s}}\Bigg[m_{B}^{4}\hat{s}\lambda^{2}+2\hat{m_{l}}^2(1+r^2+4\hat{s}-2\hat{s}^{2}+
r (-2+4\hat{s}))\Bigg]|G|^2 + \frac{1}{r} 6 m_{B}^{4}
\hat{m_{l}}^2 \hat{s} \lambda|H|^2 \nnb\\&+& \frac{1}{r\hat{s}} 2
m_{B}^{2}
(2\hat{m_{l}}^2+\hat{s})\Bigg(r^3+(-1+\hat{s})^3-r^2(3+\hat{s})-r(-3+2\hat{s}+\hat{s}^{2})\Bigg)Re(B^{*}C)\nnb\\&+&
\frac{1}{r\hat{s}}\Bigg[2
m_{B}^{2}\lambda\Bigg(2\hat{m_{l}}^2(-1+r-2\hat{s})+\hat{s}(-1+r+\hat{s})\Bigg)\Bigg]Re(F^{*}G)\nnb\\&-&
\frac{1}{r} 12 m_{B}^{2} \hat{m_{l}}^2\lambda Re(F^{*}H)-
\frac{1}{r} 12 m_{B}^{4} \hat{m_{l}}^2\lambda (-1+r) Re(G^{*}H)
\Bigg\}\eea

\section{Lepton Polarization Asymmetries \label{s4}}

Now, we would like to discuss the lepton polarizations in the $B
\rar K_1 l^+ l^- $ decays. For calculation of the double lepton
polarization asymmetries, in the rest frame of $l^{+}l^{-}$, unit
vectors $s_{i}^{\mp\mu}$ ($i=L,T,N$) are defined as
\cite{sirvanli1,Aliev2}

\bea s_L^{-\mu} &=& ( 0,\vec{e}_L^{-} ) = \left( 0,
\frac{\vec{p}_{-}}{\vel \vec{p}_{-} \ver} \right)~, \nnb \\
s_T^{-\mu} &=& ( 0,\vec{e}_T^{-} ) =   \left( 0, \vec{e}_L^{-}
\times \vec{e}_N^{-} \right)~, \nnb \\
s_N^{-\mu} &=& ( 0,\vec{e}_N^{-} ) = \left(
0,\frac{\vec{p_{K_{1}}}\times\vec{p}_{-}}{|\vec{p_{K_{1}}}\times\vec{p}_{-}|}
\right)~, \nnb \\ s_L^{+\mu} &=& ( 0,\vec{e}_L^{+} ) = \left( 0,
\frac{\vec{p}_{+}}{\vel \vec{p}_{+} \ver} \right)~, \nnb \\
s_T^{+\mu} &=& ( 0,\vec{e}_T^{+} ) =   \left( 0, \vec{e}_L^{+}
\times \vec{e}_N^{+} \right)~, \nnb \\
s_N^{+\mu} &=& ( 0,\vec{e}_N^{+} ) = \left(
0,\frac{\vec{p_{K_{1}}}\times\vec{p}_{+}}{|\vec{p_{K_{1}}}\times\vec{p}_{+}|}
\right)~. \eea

where $\vec{p}_{\pm}$ and $\vec{p}_{K_{1}}$ are the three-momenta
of the leptons $l^{+}l^{-}$ and $K_{1}$ meson in the center of
mass frame (CM) of $l^{+}l^{-}$ system, respectively. The
longitudinal unit vector $S_{L}$ is boosted to the CM frame
$l^{+}l^{-}$ under the Lorentz transformation:

\bea ( s_L^{\mp\mu} )_{CM} = ( \frac{\vel \vec{p}_\mp \ver}{m_l},
\frac{E_l \vec{p}_\mp }{m_l \vel \vec{p}_\mp \ver} )~, \eea

where $\vec{p}_{+}=-\vec{p}_{-}$, $E_{l}$ and $m_{l}$ are the
energy and mass of leptons in the CM frame, respectively. The
transversal and normal unit vectors $s_T^{\mp\mu}$, $s_N^{\mp\mu}$
are not changed under the Lorentz boost. The double lepton
polarization asymmetries are defined as:

\bea
P_{i}^{\mp}(s)=\frac{\frac{d\Gamma}{ds}(\vec{n}^{\mp}=\vec{e}_{i}^{\mp})-\frac{d\Gamma}{ds}(\vec{n}^{\mp}=-\vec{e}_{i}^{\mp})}
{\frac{d\Gamma}{ds}(\vec{n}^{\mp}=\vec{e}_{i}^{\mp})+\frac{d\Gamma}{ds}(\vec{n}^{\mp}=-\vec{e}_{i}^{\mp})}\eea

where $\vec{n}^{\mp}$ is the unit vectors in the rest frame of the
lepton. The next step, we calculated double-lepton polarization
asymmetries which is define as $P_{ij}$:

\bea \label{PLL} P_{LL}&=& \frac{1}{\Delta}\frac{2}{3}m_{B}^{2}
\Bigg\{2 m_{B}^{4} (2\hat{m_{l}}^2-\hat{s})\lambda |A|^2 +
\frac{1}{r\hat{s}}(2\hat{m_{l}}^2-\hat{s})(r^2+(-1+\hat{s})^2+2r(-1+5\hat{s}))|B|^2
\nnb\\&+&
\frac{1}{r\hat{s}}m_{B}^{4}(2\hat{m_{l}}^2-\hat{s})\lambda^{2}|C|^2+2
m_{B}^{4}(4\hat{m_{l}}^2-\hat{s})\lambda|E|^2\nnb\\&-&\frac{1}{r\hat{s}}\Bigg[\hat{s}
\Bigg(r^2+(-1+\hat{s})^2+2 r
(-1+5\hat{s})\Bigg)-2\hat{m_{l}}^2\Bigg(5r^2+5(-1+\hat{s})^2+2 r
(-5+7\hat{s}\Bigg)\Bigg]|F|^2\nnb\\&-&
\frac{1}{r\hat{s}}\Bigg[m_{B}^{4}\hat{s}\lambda^{2}-2\hat{m_{l}}^2(5+5r^2-4\hat{s}+2\hat{s}^{2}-
2r (5+2\hat{s}))\Bigg]|G|^2 + \frac{1}{r} 6 m_{B}^{4}
\hat{m_{l}}^2 \hat{s} \lambda|H|^2 \nnb\\&+& \frac{1}{r\hat{s}}
2m_{B}^{2}
(2\hat{m_{l}}^2-\hat{s})\Bigg(r^3+(-1+\hat{s})^3-r^2(3+\hat{s})-r(-3+2\hat{s}+\hat{s}^{2})\Bigg)Re(B^{*}C)\nnb\\&+&
\frac{1}{r\hat{s}}\Bigg[2
m_{B}^{2}\lambda\Bigg(2\hat{m_{l}}^2(-5+5r+2\hat{s})-\hat{s}(-1+r+\hat{s})\Bigg)\Bigg]Re(F^{*}G)\nnb\\&-&
\frac{1}{r} 12 m_{B}^{2} \hat{m_{l}}^2\lambda Re(F^{*}H)-
\frac{1}{r} 12 m_{B}^{4} \hat{m_{l}}^2\lambda (-1+r) Re(G^{*}H)
\Bigg\}\eea

\bea \label{PNN} P_{NN}&=& \frac{1}{\Delta}\frac{2}{3}m_{B}^{2}
\Bigg\{m_{B}^{4} (-4\hat{m_{l}}^2+\hat{s})\lambda |A|^2 +
\frac{1}{r\hat{s}}\Bigg(\hat{s}\lambda +
2\hat{m_{l}}^2(r^2+(-1+\hat{s})^2+2r(-1+5\hat{s}))\Bigg)|B|^2
\nnb\\&-&
\frac{1}{r\hat{s}}m_{B}^{4}(2\hat{m_{l}}^2+\hat{s})\lambda^{2}|C|^2+
m_{B}^{4}(4\hat{m_{l}}^2-\hat{s})\lambda|E|^2+
\frac{1}{r\hat{s}}(2\hat{m_{l}}^2+\hat{s})\lambda|F|^2 \nnb\\&+&
\frac{1}{r\hat{s}}\Bigg[m_{B}^{4}\hat{s}\lambda^{2}+2\hat{m_{l}}^2(1+r^2+4\hat{s}-2\hat{s}^{2}+
2r(-2+4\hat{s}))\Bigg]|G|^2 + \frac{1}{r} 6 m_{B}^{4}
\hat{m_{l}}^2 \hat{s}\lambda|H|^2 \nnb\\&-& \frac{1}{r\hat{s}}
2m_{B}^{2}
(2\hat{m_{l}}^2+\hat{s})\Bigg(r^3+(-1+\hat{s})^3-r^2(3+\hat{s})-r(-3+2\hat{s}+\hat{s}^{2})\Bigg)Re(B^{*}C)\nnb\\&+&
\frac{1}{r\hat{s}}\Bigg[2
m_{B}^{2}\lambda\Bigg(2\hat{m_{l}}^2(-1+r-2\hat{s})+\hat{s}(-1+r+\hat{s})\Bigg)\Bigg]Re(F^{*}G)\nnb\\&-&
\frac{1}{r} 12 m_{B}^{2} \hat{m_{l}}^2\lambda Re(F^{*}H)-
\frac{1}{r} 12 m_{B}^{4} \hat{m_{l}}^2\lambda (-1+r) Re(G^{*}H)
\Bigg\}\eea

\bea \label{PTT} P_{TT}&=& \frac{1}{\Delta}\frac{2}{3}m_{B}^{2}
\Bigg\{m_{B}^{4} (4\hat{m_{l}}^2+\hat{s})\lambda |A|^2 +
\frac{1}{r\hat{s}}\Bigg(-\hat{s}\lambda+2\hat{m_{l}}^2(r^2+(-1+\hat{s})^2+2r(-1+5\hat{s}))\Bigg)|B|^2
\nnb\\&+&
\frac{1}{r\hat{s}}m_{B}^{4}(2\hat{m_{l}}^2-\hat{s})\lambda^{2}|C|^2+
m_{B}^{4}(4\hat{m_{l}}^2-\hat{s})\lambda|E|^2\nnb\\&+&\frac{1}{r\hat{s}}\lambda
(-10m_{l}^2+\hat{s})|F|^2\nnb\\&+&
\frac{1}{r\hat{s}}\Bigg[m_{B}^{4}\hat{s}\lambda^{2}-2\hat{m_{l}}^2(5+5r^2-4\hat{s}+2\hat{s}^{2}-
2r (5+2\hat{s}))\Bigg]|G|^2 - \frac{1}{r} 6 m_{B}^{4}
\hat{m_{l}}^2 \hat{s} \lambda|H|^2 \nnb\\&+& \frac{1}{r\hat{s}}
2m_{B}^{2}
(2\hat{m_{l}}^2-\hat{s})\Bigg(r^3+(-1+\hat{s})^3-r^2(3+\hat{s})-r(-3+2\hat{s}+\hat{s}^{2})\Bigg)Re(B^{*}C)\nnb\\&+&
\frac{1}{r\hat{s}}\Bigg[2
m_{B}^{2}\lambda\Bigg(-2\hat{m_{l}}^2(-5+5r+2\hat{s})+\hat{s}(-1+r+\hat{s})\Bigg)\Bigg]Re(F^{*}G)\nnb\\&+&
\frac{1}{r} 12 m_{B}^{2} \hat{m_{l}}^2\lambda Re(F^{*}H)+
\frac{1}{r} 12 m_{B}^{4} \hat{m_{l}}^2\lambda (-1+r) Re(G^{*}H)
\Bigg\}\\
P_{LN}&=&
\frac{1}{\Delta}\frac{1}{r\sqrt{\hat{s}}}m_{B}^{2}\hat{m_{l}}\pi
\sqrt{\lambda}\Bigg[(-1+r+\hat{s})Im(B^{*}F)+m_{B}^{2}\lambda
Im(C^{*}F) \nnb\\&+& m_{B}^{2}(-1+r)(-1+r+\hat{s})Im(B^{*}G)+
m_{B}^{4}(-1+r)\lambda
Im(C^{*}G)\nnb\\&-&m_{B}^{2}\hat{s}(-1+r+\hat{s})Im(B^{*}H)-m_{B}^{4}\hat{s}\lambda
Im(C^{*}H)\Bigg]\\
P_{LT}&=&
\frac{1}{\Delta}\frac{1}{\sqrt{\hat{s}}}m_{B}^{2}\hat{m_{l}}\pi
\lambda \sqrt{1-\frac{4\hat{m_{l}}^2}{\hat{s}}}
\Bigg(-\frac{1}{r}(-1+r+\hat{s})|F|^2-\frac{1}{r}m_{B}^{4}(-1+r)\lambda
|G|^2+2m_{B}^{2}\hat{s}Re(B^{*}E)\nnb\\&+&
2m_{B}^{2}\hat{s}Re(A^{*}F)-\frac{1}{r}m_{B}^{2}(2+2r^2+\hat{s}^2-r(4+\hat{s}))Re(F^{*}G)+
\frac{1}{r}m_{B}^{2}\hat{s}(-1+r+\hat{s})Re(F^{*}H)\nnb\\&+& \frac{1}{r}m_{B}^{4}\hat{s}\lambda Re(G^{*}H)\Bigg)\\
P_{TL}&=& \frac{1}{\Delta}
\frac{1}{\sqrt{\hat{s}}}m_{B}^{2}\hat{m_{l}}\pi \lambda
\sqrt{1-\frac{4\hat{m_{l}}^2}{\hat{s}}}
\Bigg(-\frac{1}{r}(-1+r+\hat{s})|F|^2-\frac{1}{r}m_{B}^{4}(-1+r)\lambda
|G|^2-2m_{B}^{2}\hat{s}Re(B^{*}E)\nnb\\&-&
2m_{B}^{2}\hat{s}Re(A^{*}F)-\frac{1}{r}m_{B}^{2}(2+2r^2-3\hat{s}+\hat{s}^2-r(4+\hat{s}))Re(F^{*}G)+
\frac{1}{r}m_{B}^{2}\hat{s}(-1+r+\hat{s})Re(F^{*}H)\nnb\\&+& \frac{1}{r}m_{B}^{4}\hat{s}\lambda Re(G^{*}H)\Bigg)\\
P_{TN}&=& \frac{1}{\Delta}\frac{-4m_{B}^{2}\lambda}{3r}\sqrt{1-\frac{4\hat{m_{l}}^2}{\hat{s}}}\Bigg[-m_{B}^{4}r\hat{s} Im(A^{*}E)+
Im(B^{*}F)+m_{B}^{2}\Bigg((-1+r+\hat{s})Im(C^{*}F)\nnb\\&+&(-1+r+\hat{s})Im(B^{*}G)+m_{B}^{2}\lambda Im(C^{*}G)\Bigg)\Bigg]\\
P_{NL}&=&
\frac{1}{\Delta}\frac{1}{r\sqrt{\hat{s}}}m_{B}^{2}\hat{m_{l}}\pi
\lambda \Bigg[-(-1+r+\hat{s})Im(B^{*}F)-m_{B}^{2}\lambda
Im(C^{*}F)\nnb\\&-& m_{B}^{2}(-1+r)(-1+r+\hat{s})Im(B^{*}G)-
m_{B}^{4}(-1+r)\lambda Im(C^{*}G)\nnb\\&+& m_{B}^{2}\hat{s}(-1+r+\hat{s})Im(B^{*}H)+m_{B}^{2}\hat{s}\lambda Im(C^{*}H)\Bigg] \\
P_{NT}&=&
\frac{1}{\Delta}\frac{4m_{B}^{2}\lambda}{3r}\sqrt{1-\frac{4\hat{m_{l}}^2}{\hat{s}}}\Bigg[-m_{B}^{4}
r
\hat{s}Im(A^{*}E)+Im(B^{*}F)+m_{B}^{2}\Bigg((-1+r+\hat{s})Im(C^{*}F)\nnb\\&+&(-1+r+\hat{s})Im(B^{*}G)+m_{B}^{2}\lambda
Im(C^{*}G)\Bigg)\Bigg]\eea

\section{Numerical analysis and discussion \label{s5}}
In this section, we present our numerical results on the double
lepton polarization asymmetries for the $B \rar K_1 l^+ l^- $
decays. First, we present the values of input parameters are:

\begin{eqnarray} & &
m_{B} =5.28\, GeV \, , \, m_{B^{*}} =5.32 \, GeV \, ,  m_{K_{1}} =1.402\, GeV \, , \, \nnb \\
& &
m_{b} =4.8 \, GeV \, , m_{s} =0.13 \, GeV \, , m_{\mu} =0.105 \, GeV \, , m_{\tau} =1.77 \, GeV \, ,  \nnb \\
& & |V_{tb} V^*_{ts}|=0.04 \, \, , \, \, \alpha^{-1}=137\, , \, \,
G_F=1.17 \times 10^{-5}\, GeV^{-2} \,  ,\,\, \tau_{B}=1.53 \times
10^{-12} \, s .
\end{eqnarray}

The $B\rar K$ transition form factors are the main input
parameters in performing the numerical analysis, which are
embedded into the expressions of the double-lepton polarization
asymmetries. For them we have used their expression given by Eq.
(8-15). The differential decay rate for $B \rar K_1 l^+ l^- $ can
be defined in terms of integration on $\hat{s}$, which is
determined to the range of the $4\hat{m_{l}}^2\leq
s\leq(m_{B}-m_{K_{1}})^2$.

In Fig.1, we present the dependence of the $P_{LL}$ for the $B
\rar K_1 \mu^+ \mu^- $ decay as a
    function of $s/m_{B}^{2}$. We see that, $P_{LL}$ in UED
    compatible with the SM result. Increasing $\hat{s}$, $P_{LL}$
    is moderate for the low of $\hat{s}$. The effect of KK contribution in the Wilson
    coefficient are consistent for $1/R=200 GeV$ at low value of
    $\hat{s}$. $1/R=200 GeV$ value is greater than $1/R=400 GeV$.
In Fig.2, Double lepton longitudinal polarization asymmetries for
the $B \rar K_1 \tau^+ \tau^- $ decay is presented,
    From this figure is follows, UED model prediction coincide with the SM
    result. One can see that the value of the longitudinal
    polarization is different in the low of $\hat{s}$ for the $B \rar K_1 \tau^+ \tau^- $
    decay. While $1/R=200 GeV$ value is max in the UED model, The
    SM result is approximately two times lower than this value.
In Fig.3, For the $B \rar K_1 \mu^+ \mu^- $ decay, we
    analysis to the normal polarizations. We obtained good result
    at the $1/R=200 GeV$ in UED model. We can see that the effect of extra dimension are very
    noticeable at the small value of $\hat{s}$. When the value of
    $\hat{s}$ close to $0.2$, all the value of normal polarization
    is coincide with each other. In $\hat{s}=0.36$, the value of $1/R=200 GeV$ is five
    times bigger than SM result.
    But in Fig.4, for the $B \rar K_1 \tau^+ \tau^- $
    decay, it is similar to the $P_{LL}$ result.
In Fig.5, We examine to the transversal polarization for
    the $B \rar K_1 \mu^+ \mu^- $ decay. At the $1/R=200 GeV$ value, we compared to that
    of the SM prediction $P_{TT}$ is larger from SM. Again, the
    effects of extra dimension are distinguished at the small
    value of momentum transfer $\hat{s}$ where $P_{TT}$ is
    minimum. For the $\hat{s}=0.53$ value, all polarization values are
    decreases.
In Fig.6, We analysis to transversal polarization as a function of
the $\hat{s}$ for the $B \rar K_1 \tau^+ \tau^- $ decay. We
observe a little contributions from UED model, especially in the
$1/R=400 GeV$ value. But UED model is better than SM in this
figure. All model values come together with the SM result in the
$\hat{s}=0.53$ value. In Fig.7, we investigate $P_{LT}$
polarization. We see that increasing $\hat{s}$, $P_{LT}$ increase
until $\hat{s}=0.5 GeV^2$. After this value of $\hat{s}$ two
models are decrease until $\hat{s}=0.55 GeV^2$.
$(P_{LT})_{UED}=2(P_{LT})_{SM}$ at $1/R=200 GeV$. So It is also
very useful for establishing new physics.
In Fig.8, We show our
predictions for the $P_{TL}$ for $B \rar K_1 \tau^+ \tau^- $
    decay. We get $|(P_{TL})_{UED}|>|(P_{TL})_{SM}|$. This result
    can serve as a good test for discrimination of two models.
   The other polarizations for the $B \rar K_1 l^+ l^- $
    decay, we have imaginary part and therefore there is no
    interference terms between SM and UED model contributions.

In conclusion, we have studied the double-lepton polarization
asymmetries in the UED model. We obtain different double-lepton
polarization asymmetries which is very sensitive to the UED model.
It has been shown that all these physical observebles are very
sensitive to the existence of new physics beyond SM and their
experimental measurements can give valuable information on it.

\vspace{2cm}

\textbf{Acknowledgements}

The author would like to thank T. M. Aliev, M. Savc{\i} and A.
Ozpineci for useful discussions during the course of the work.

\newpage
\renewcommand{\topfraction}{.99}
\renewcommand{\bottomfraction}{.99}
\renewcommand{\textfraction}{.01}
\renewcommand{\floatpagefraction}{.99}

\begin{figure}
\centering
\includegraphics[width=5in]{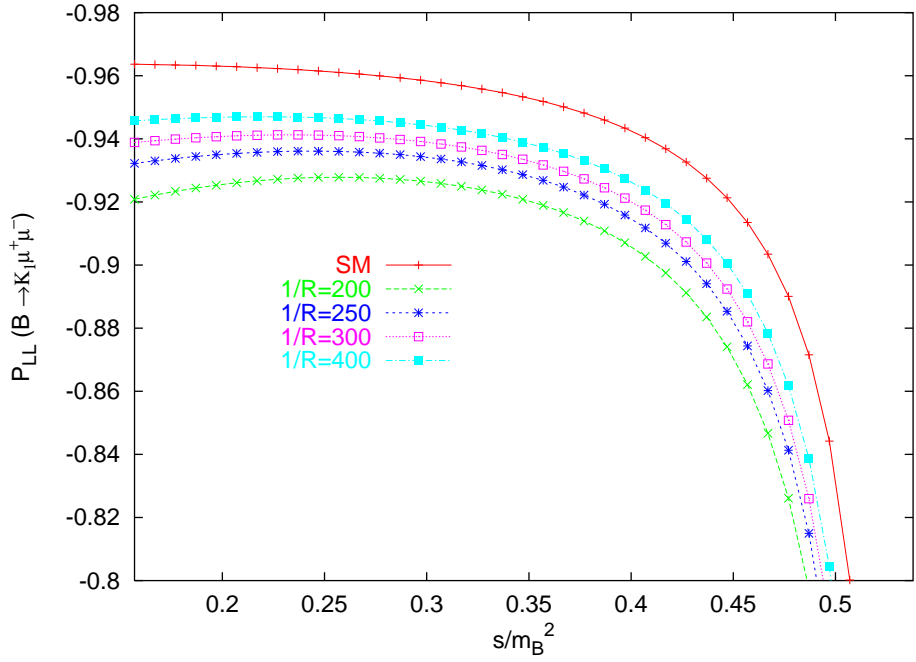}
\caption{The dependence of the Longitudinal polarization,for $B \rar K_1 \mu^+ \mu^- $ decay, as a
function of the $\hat{s}$ \label{f1}.}
\end{figure}
\begin{figure}
\centering
\includegraphics[width=5in]{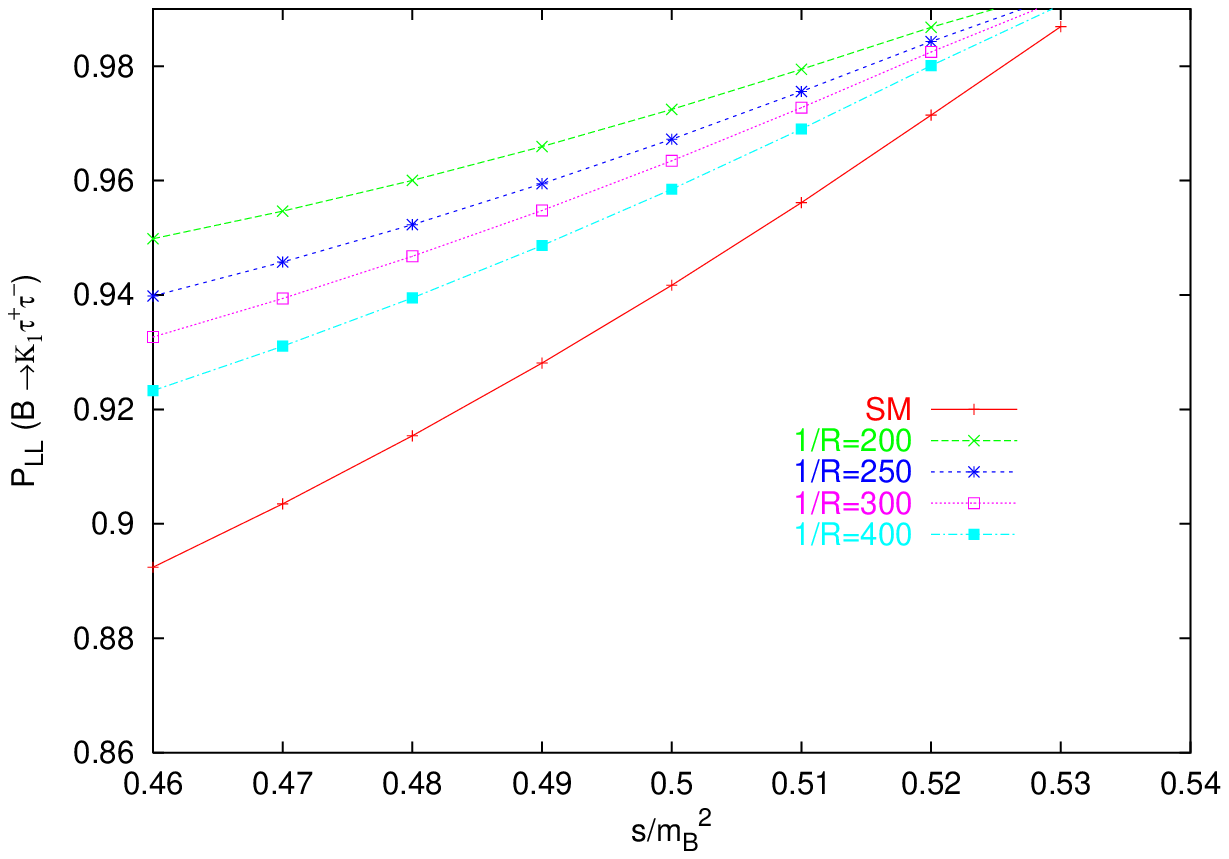}
\caption{The dependence of the Longitudinal polarization,for $B \rar K_1 \tau^+ \tau^- $ decay, as a
function of the $\hat{s}$ . \label{f2}.}
\end{figure}
\clearpage
\begin{figure}
\centering
\includegraphics[width=5in]{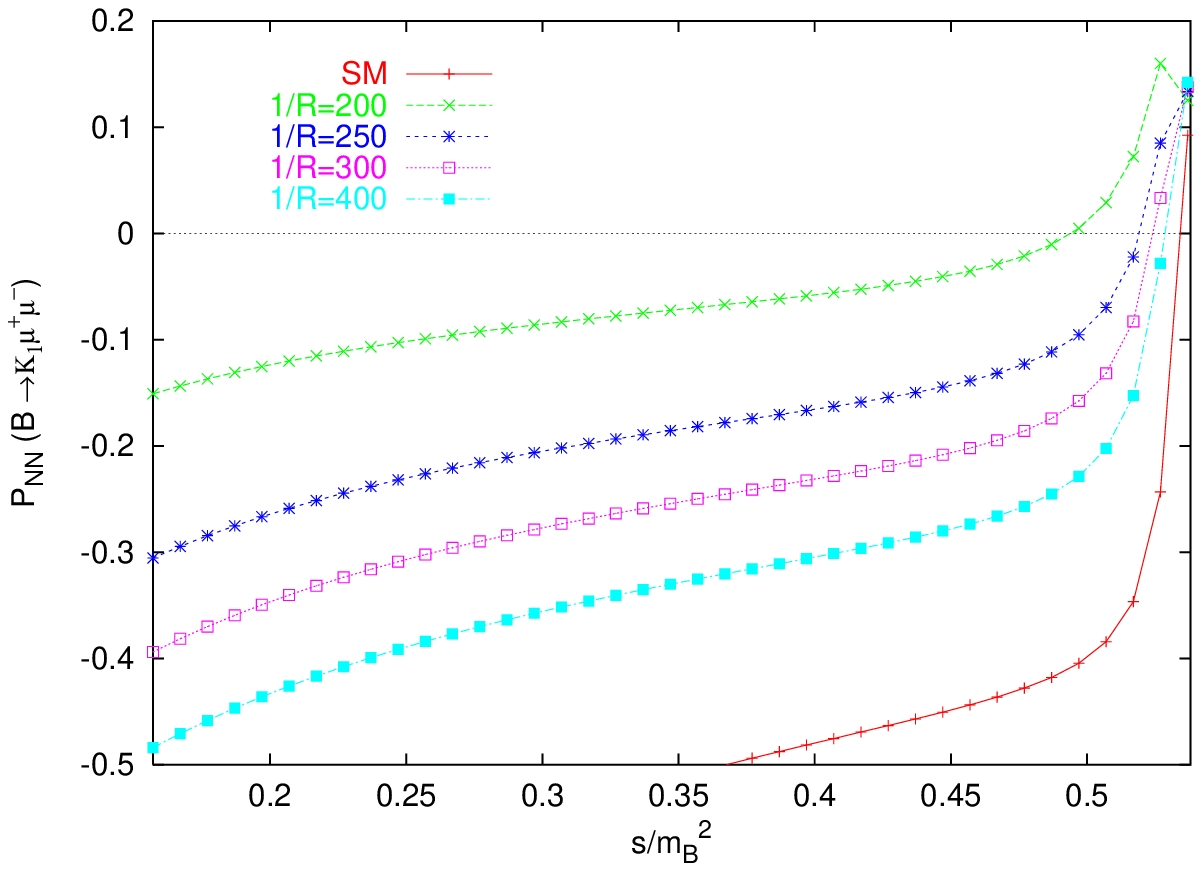}
\caption{The dependence of the Normal polarization,for $B \rar K_1
\mu^+ \mu^- $ decay, as a function of the $\hat{s}$ .\label{f3}}
\end{figure}
\begin{figure}
\centering
\includegraphics[width=5in]{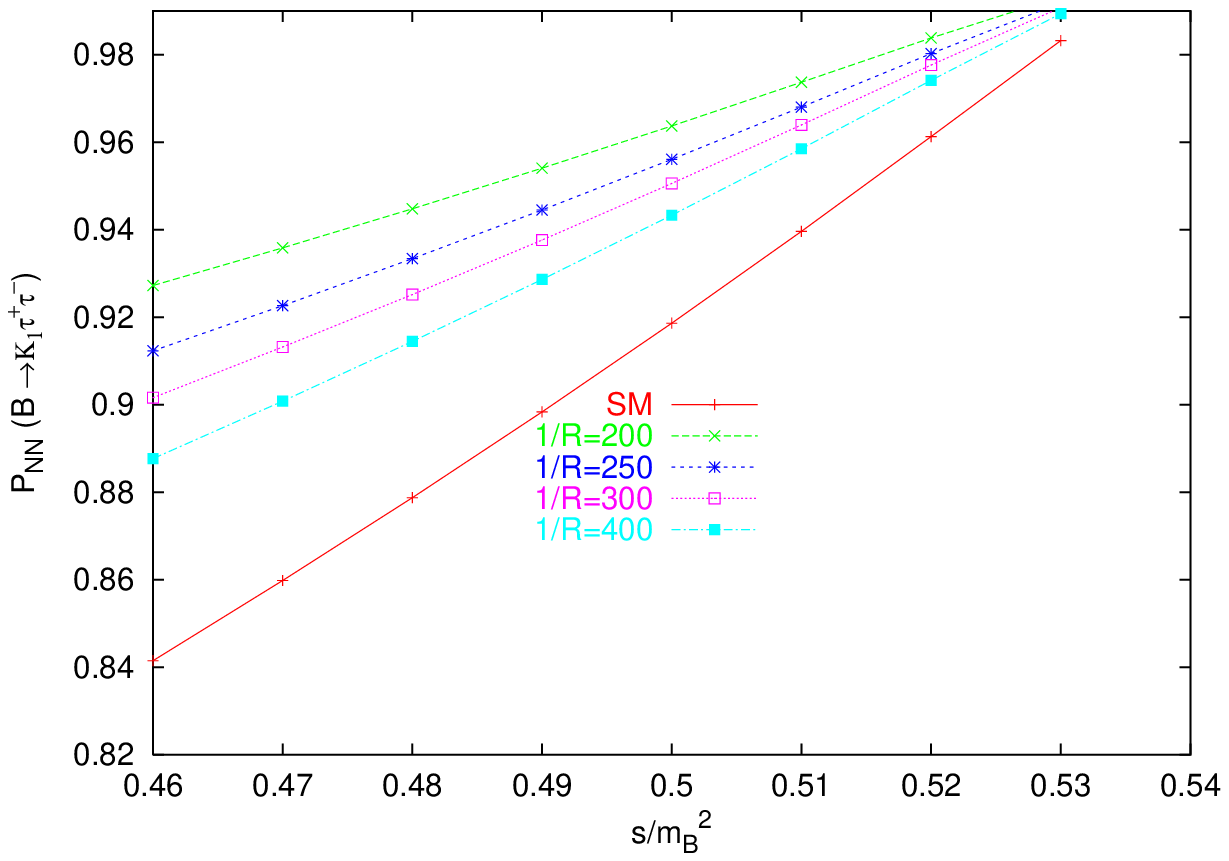}
\caption{The dependence of the Normal polarization,for $B \rar K_1 \tau^+ \tau^- $ decay, as a
function of the $\hat{s}$ . \label{f4}}
\end{figure}

\begin{figure}
\centering
\includegraphics[width=5in]{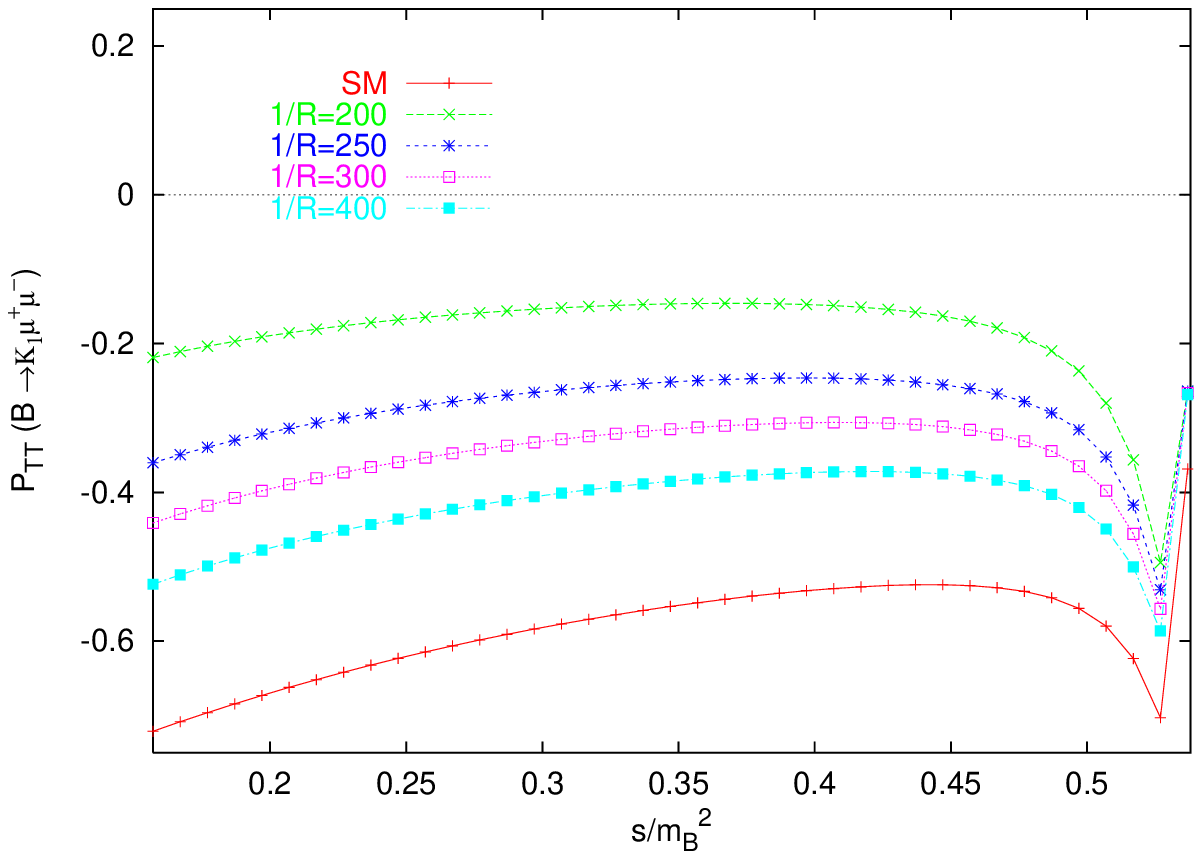}
\caption{The dependence of the Transversal polarization,for $B \rar K_1 \mu^+ \mu^- $ decay, as a
function of the $\hat{s}$ . \label{f5}}
\end{figure}

\begin{figure}
\centering
\includegraphics[width=5in]{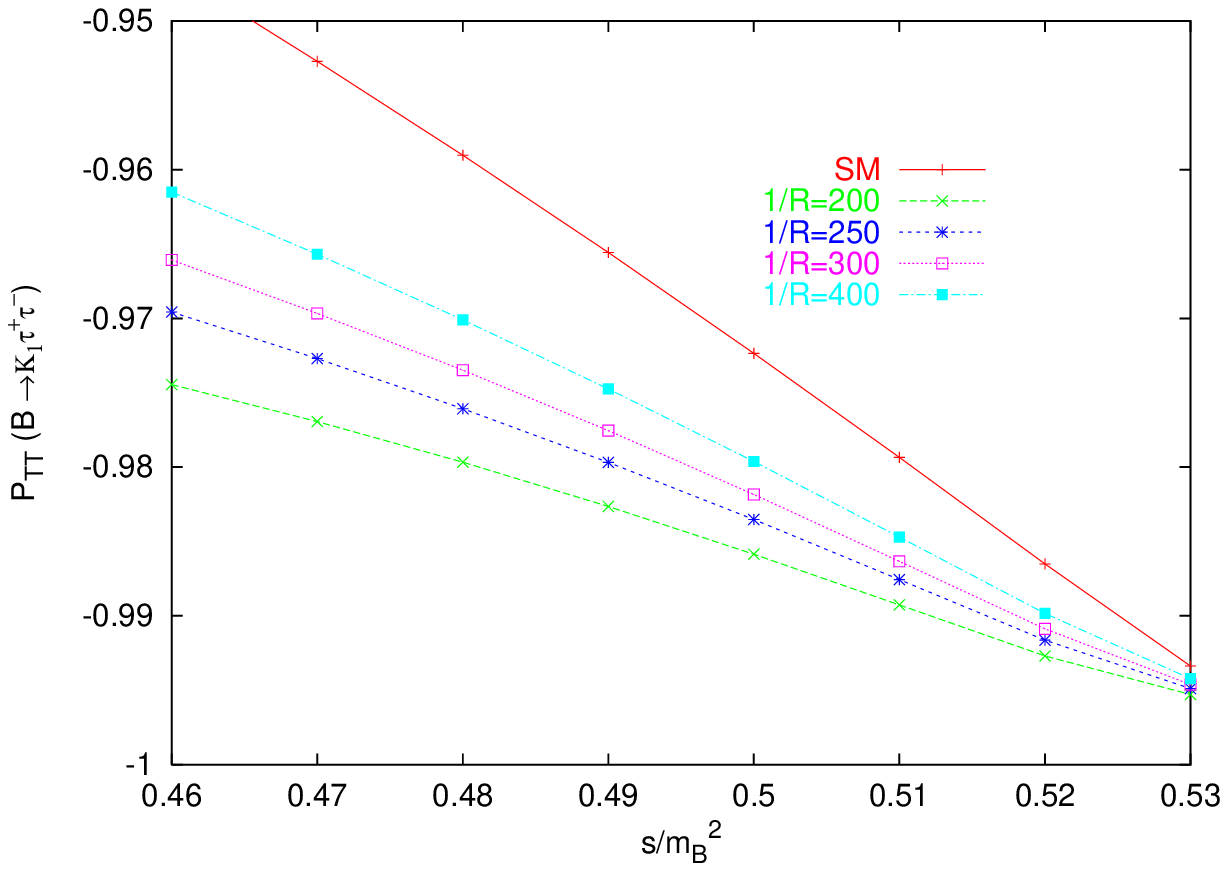}
\caption{The dependence of the Transversal polarization,for $B \rar K_1 \tau^+ \tau^- $ decay, as a
function of the $\hat{s}$ . \label{f6}}
\end{figure}

\begin{figure}
\centering
\includegraphics[width=5in]{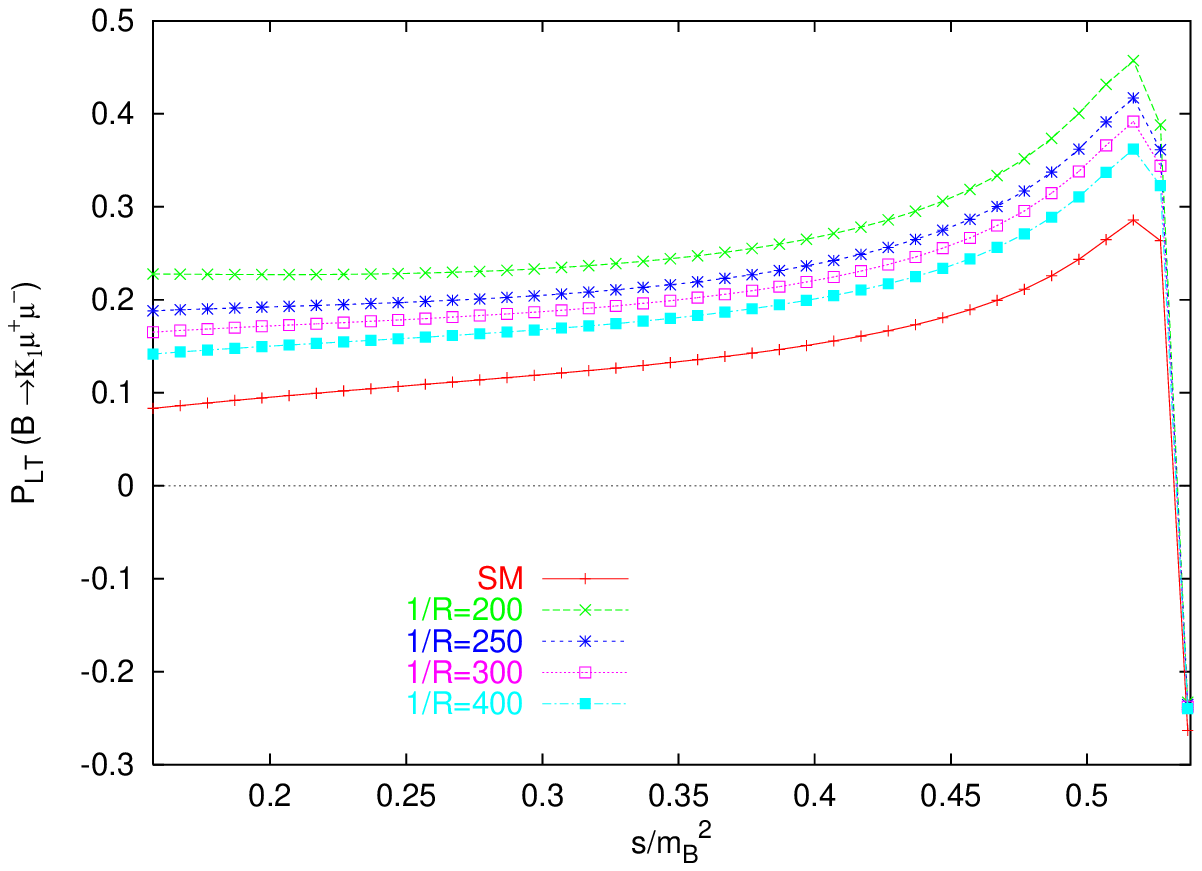}
\caption{The dependence of the $P_{LT}$ polarization,for $B \rar K_1 \mu^+ \mu^- $ decay, as a
function of the $\hat{s}$ . \label{f7}}
\end{figure}

\begin{figure}
\centering
\includegraphics[width=5in]{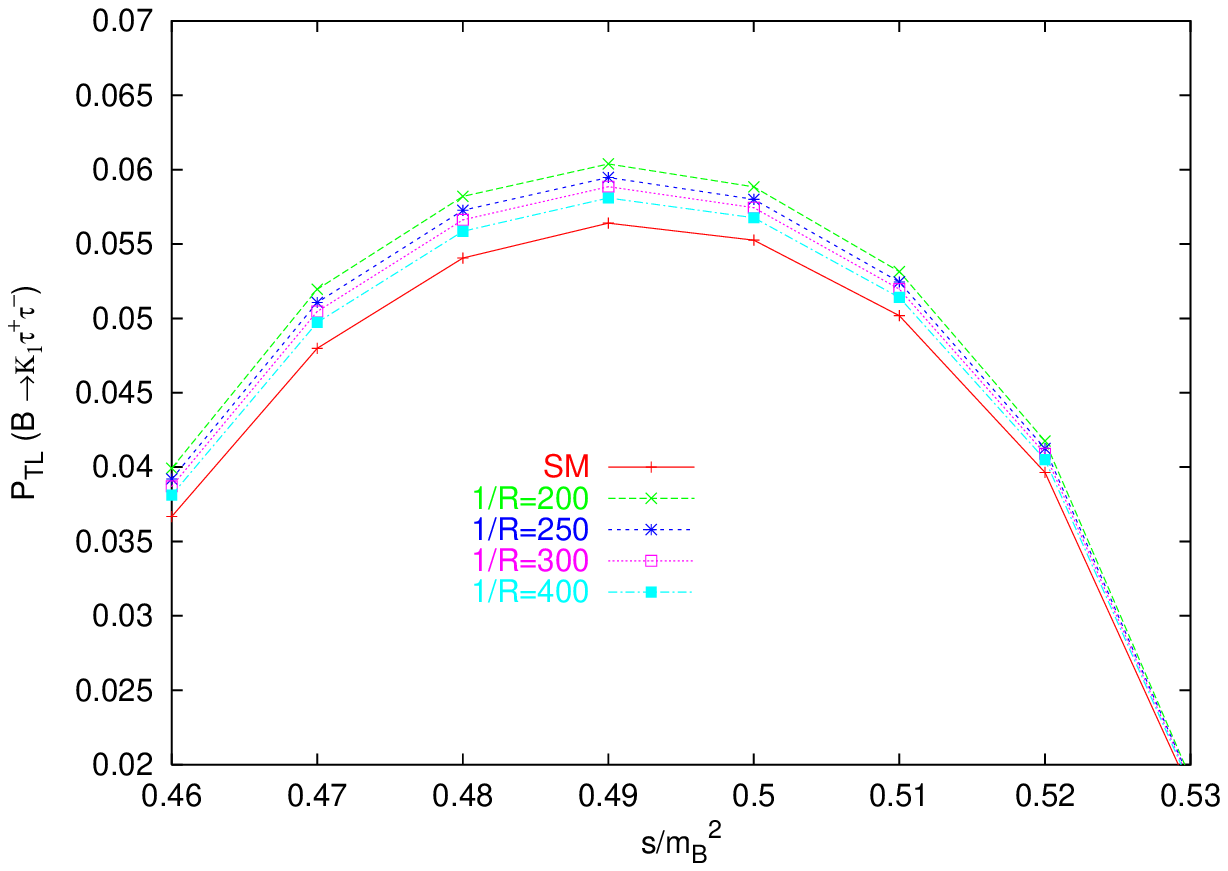}
\caption{The dependence of the $P_{TL}$ polarization,for $B \rar K_1 \tau^+ \tau^- $ decay, as a
function of the $\hat{s}$ . \label{f8}}
\end{figure}

\clearpage
%

%
%
\end{document}